\documentclass[prl,superscriptaddress,twocolumn,floatfix,showpacs,amsmath,amssymb]{revtex4}

\usepackage{graphicx}
\usepackage{dcolumn}
\usepackage{bm}
\usepackage[latin1]{inputenc}
\usepackage{graphicx,color,epsfig,rotate}

\newcommand{\lao}{$\rm LaFeAsO$}

\newcommand{\tn}{$T_{\rm{N}}$}
\newcommand{\tsdw}{$T_{\rm{N}}$}
\newcommand{\ts}{$T_{\rm{S}}$}

\begin{document}

\title{Commensurate Spin Density Wave in LaFeAsO: A Local Probe Study}


\author{H.-H.~Klauss}\email{h.klauss@physik.tu-dresden.de}
\affiliation{Institut f\"ur Festk\"orperphysik, TU Dresden, D-01069
Dresden, Germany}
\author{H.~Luetkens}\affiliation{Laboratory for Muon-Spin Spectroscopy, Paul Scherrer
Institut, CH-5232 Villigen PSI, Switzerland}
\author{R.~Klingeler}
\affiliation{Leibniz-Institut f\"ur Festk\"orper- und
Werkstoffforschung (IFW) Dresden, D-01171 Dresden, Germany}
\author{C.~Hess}
\affiliation{Leibniz-Institut f\"ur Festk\"orper- und
Werkstoffforschung (IFW) Dresden, D-01171 Dresden, Germany}
\author{F.J.~Litterst} \affiliation{Institut f\"ur Physik der Kondensierten Materie, TU Braunschweig, D-38106 Braunschweig,
Germany}
\author{M.~Kraken}\affiliation{Institut f\"ur Physik der Kondensierten Materie, TU Braunschweig, D-38106 Braunschweig,
Germany}
\author {M.M.~Korshunov}\affiliation {Max-Planck-Institut f\"ur Physik Komplexer Systeme,
D-01187 Dresden, Germany}
 \affiliation{L.V. Kirensky Institute of Physics, Siberian
Branch of Russian Academy of Sciences, 660036 Krasnoyarsk, Russia}
\author {I.~Eremin}
\affiliation {Max-Planck-Institut f\"ur Physik Komplexer Systeme,
D-01187 Dresden, Germany} \affiliation {Institut f\"ur Mathematische
Physik, TU Braunschweig, D-38106 Braunschweig, Germany}
\author {S.-L. Drechsler}
\affiliation{Leibniz-Institut f\"ur Festk\"orper- und
Werkstoffforschung (IFW) Dresden, D-01171 Dresden, Germany}
\author{R.~Khasanov}\affiliation{Laboratory for Muon-Spin Spectroscopy, Paul Scherrer Institut, CH-5232 Villigen PSI,
Switzerland}
\author{A.~Amato} \affiliation{Laboratory for Muon-Spin Spectroscopy, Paul Scherrer Institut, CH-5232 Villigen PSI,
Switzerland}
\author{J.~Hamann-Borrero}
\affiliation{Leibniz-Institut f\"ur Festk\"orper- und
Werkstoffforschung (IFW) Dresden, D-01171 Dresden, Germany}
\author{N.~Leps}
\affiliation{Leibniz-Institut f\"ur Festk\"orper- und
Werkstoffforschung (IFW) Dresden, D-01171 Dresden, Germany}
\author{A.~Kondrat}
\affiliation{Leibniz-Institut f\"ur Festk\"orper- und
Werkstoffforschung (IFW) Dresden, D-01171 Dresden, Germany}
\author{G.~Behr}
\affiliation{Leibniz-Institut f\"ur Festk\"orper- und
Werkstoffforschung (IFW) Dresden, D-01171 Dresden, Germany}
\author{J.~Werner}
\affiliation{Leibniz-Institut f\"ur Festk\"orper- und
Werkstoffforschung (IFW) Dresden, D-01171 Dresden, Germany}
\author{B.~B\"uchner}
\affiliation{Leibniz-Institut f\"ur Festk\"orper- und
Werkstoffforschung (IFW) Dresden, D-01171 Dresden, Germany}


\date{\today}

\begin{abstract}
We present a detailed study on the magnetic order in the undoped
mother compound LaFeAsO of the recently discovered Fe-based
superconductor LaFeAsO$_{1-x}$F$_x$. In particular, we present local
probe measurements of the magnetic properties of LaFeAsO by means of
$^{57}$Fe M\"ossbauer spectroscopy and muon spin relaxation in zero
external field along with magnetization and resistivity studies.
These experiments prove a commensurate static magnetic order with a
strongly reduced ordered moment of 0.25(5) $\mu_B$ at the iron site
below \tsdw\ = 138~K, well separated from a structural phase
transition at \ts\ = 156~K. The temperature dependence of the
sublattice magnetization is determined and compared to theory. Using
a four-band spin density wave model both, the size of the order
parameter and the quick saturation below \tsdw\ are reproduced.
\end{abstract}

\pacs{76.75.+i, 76.80.+y, 75.30.Fv, 74.70.-b}


\maketitle



The recently discovered Fe-based superconductors
LaFeAsO$_{1-x}$F$_x$ \cite{Kamihara08} and the related materials in
which La is substituted by Sm, Ce, Nd, Pr, and Gd, respectively
\cite{Chen08_XH-arXiv,Chen08_GFb-arXiv,Ren08a-arXiv,Ren08b-arXiv,Cheng08-arXiv,Ren08c-arXiv}
has triggered an intense research in the oxypnictides. Besides the
high critical temperature above 50~K there are further striking
similarities to the properties of the high-T$\rm_C$ cuprates. The
oxypnictides have a layered crystal structure with alternating FeAs
and LaO sheets, where the Fe atoms are arranged on a simple square
lattice \cite{Kamihara08}. Theoretical studies reveal a
two-dimensional electronic structure \cite{Singh08-arXiv} and it is
believed that conductivity takes place mainly in the FeAs layers
while the LaO layers provide the charge reservoir when doped with F
ions. Again similar as in the cuprates, superconductivity emerges
when doping a magnetic mother compound with electrons or holes and
thereby supressing the magnetic order~\cite{Luetkens08}. This
suggests an interesting interplay between magnetism and
superconductivity and, indeed, a recent theoretical work suggests
that magnetic fluctuations associated with quantum critical point
are essential for superconductivity in the electron doped
LaFeAsO$_{1-x}$F$_x$ superconductors \cite{Giovannetti08-arXiv}.

However, in contrast to the cuprates, the magnetic mother compound
is not a Mott-Hubbard insulator but a poor metal. A large covalency
in the FeAs layers was found \cite{Singh08-arXiv,Haule08-arXiv},
which in the case of tetragonal LaFePO, i.e. the compound where As
is replaced by P, leads to a non-magnetic ground state
\cite{Lebegue07,Liang07}. In contrast, in LaFeAsO there is an
additional structural distortion at elevated temperatures
\cite{Cruz08-arXiv,Nomura08} and a long range spin density wave
(SDW) antiferromagnetic order has been observed in neutron
scattering experiments on powder samples below
$\sim$150~K.\cite{Cruz08-arXiv}. First principle calculations yield
antiferromagnetic order with Fe magnetic moments ranging from
1.5~$\mu_\mathrm{B}$ to 2.3~$\mu_\mathrm{B}$
\cite{Ma08-arXiv,Dong08-arXiv,Cao08-arXiv,Giovannetti08-arXiv},
while the neutron scattering experiments indicate a much smaller
value. Assuming that the full sample volume is contributing to the
magnetic scattering an ordered moment of $\sim$
0.35~$\mu_\mathrm{B}$ \cite{Cruz08-arXiv} is inferred from the weak
superlattice reflections in powder neutron diffraction. A local
probe measurement, which could verify the type of order and the size
of the ordered moment in LaOFeAs, is up to now lacking. It is,
however, apparent that a detailed knowledge and understanding of the
magnetic properties of LaOFeAs form the basis to tackle the
intriguing question about the interplay between magnetism and
superconductivity in this new class of superconductors.

In this Letter we report local probe measurements of the magnetic
properties of LaOFeAs by means of $^{57}$Fe M\"ossbauer spectroscopy
and muon spin relaxation, both in zero external field. These studies
prove a static magnetic order below \tsdw\ = 138~K with a clearly
commensurate spin structure and a strongly reduced ordered moment at
the Fe site in the ordered phase. The data provide a high precision
measurement of the temperature dependence of the sublattice
magnetization which is in fair agreement with a theoretical model
assuming a four-band SDW model. The theoretical calculations
reproduce the size of the order parameter as well as the quick
saturation below \tsdw , which markedly differs from the
conventional mean field behavior.

Polycrystalline \lao\ has been prepared by using a two-step solid
state reaction method, similar to that described by Zhu et al.
\cite{Zhu_X08-arXiv}, and annealed in vacuum.
The crystal structure and the composition were investigated by
powder X-ray diffraction and wavelength-dispersive X-ray
spectroscopy (WDX). From the X-ray diffraction data impurity
concentrations  smaller than 1~\% are inferred. In order to
investigate the magnetic order we have performed zero field $\mu$SR
between 1.6 and 300~K. $^{57}$Fe M\"ossbauer spectroscopy
experiments have been done in the temperature range 13~K to 180~K
$[$source: $^{57}$Co-in-Rh matrix at room temperature; emission line
half width at half maximum: 0.130(2) mm/s$]$. The results of these
local probe experiments are compared with magnetization and
resistivity data.

\begin{figure}[htbp]
\center{\includegraphics[width=0.9\columnwidth,angle=0,clip]{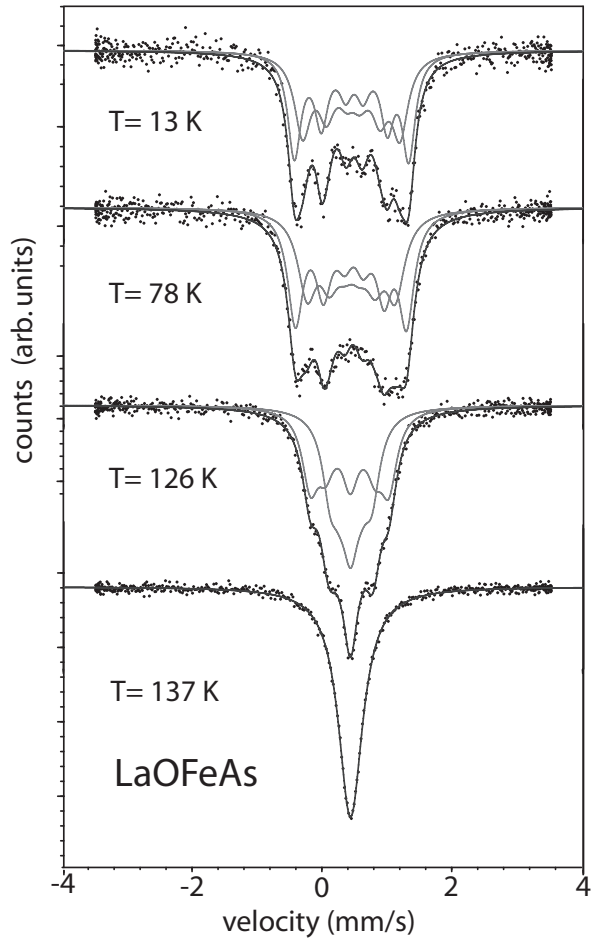}}
    \caption[]{Typical M\"ossbauer spectra of a LaFeAsO powder sample.
    The full lines describe the fit and the subspectra used in the data analysis (see text)}
\label{moessbauer-spectra}
\end{figure}

Typical M\"ossbauer spectra are shown in figure
\ref{moessbauer-spectra}. Above 140~K, the spectra can be fitted
with a single Lorentzian line
 with an isomer shift of $S$=0.52(1) mm/s representing \lao . This
is in the typical range of low or intermediate spin Fe(II). Below
140~K, a splitting of the absorption line reveals the formation of a
magnetic hyperfine field. In this temperature range the spectra have
been analyzed by diagonalizing the hyperfine Hamiltonian including
electric quadrupole and magnetic hyperfine interaction. The main
result is the observation of a magnetic hyperfine field with a mean
saturation value at low temperatures of $B_{hf}(0)$= 4.86(5)~T. The
data are consistent with a commensurate antiferromagnetic order of
the iron spins. An incommensurate spin density wave can be ruled out
since this would lead to a very broad hyperfine field distribution
ranging from zero to a maximum field value. In our experiment  the
hyperfine field distribution is broadenend only slightly indicating
a slightly inhomogeneous magnetic state in our sample, which could
be caused by small variations of the oxygen content. In order to
properly describe the spectra we have used two sextets of equal
strength which technically account for the broadening. Both sextets
show an electric quadrupole splitting of $QS \approx$ 0.3 mm/s due
to a small deformation of the FeAs$_4$ tetrahedron below the
structural phase transition \cite{Cruz08-arXiv}.

The second important information from the M\"ossbauer experiments
concerns the size of the ordered moment. The hyperfine field of
4.86~T is about one order of magnitude smaller than the observed
hyperfine field for iron oxide compounds which exhibit a fully
ordered moment of 2~$\mu_B$ \cite{Greenwood-71}. This strongly
contradicts all models for the magnetism in \lao\ based on local
high spin Fe moments. The ordered moment  at the Fe site in LaFeAsO
can be estimated to 0.25(5) $\mu_B$. The error bar in this
determination mainly reflects the not well known influence of
covalency, i.e. a possible delocalization of spin density from the
Fe 3d to adjacent As atoms. It is well established that a strong
covalency could, in principle, reduce the measured hyperfine field
at the iron nucleus as determined by M\"ossbauer spectroscopy.
However, electronic structure calculations reveal only a very small
contribution from As orbitals to the electronic bands close to the
Fermi level \cite{Mazin08-arXiv} suggesting that the influence of
covalency on $B_{hf}(0)$ is weak for LaFeAsO.

\begin{figure}[htbp]
\center{\includegraphics[width=0.95\columnwidth,angle=0,clip]{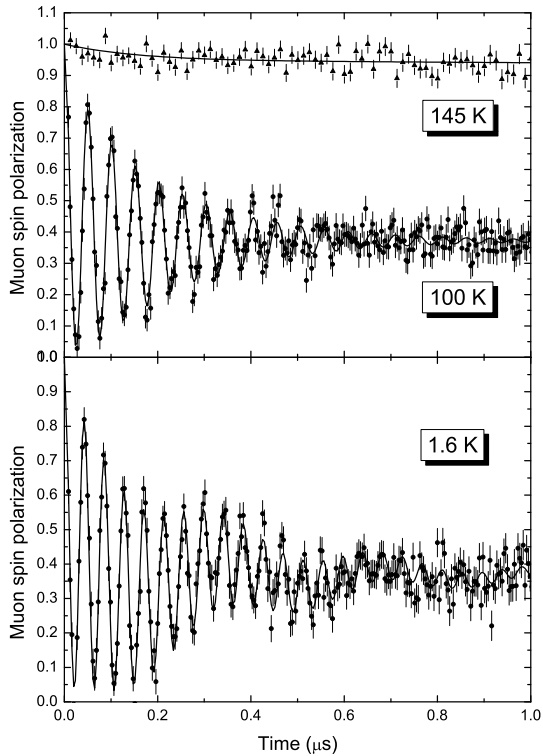}}
\caption[]{Zero field $\mu$SR spectra of LaFeAsO for 1.6, 100, and
145~K.} \label{ZF-spectra}
\end{figure}

Further insight into the magnetic order is obtained from our $\mu$SR
data. Representative spectra of \lao\ for  temperatures above and
below \tsdw\ are shown in fig.~\ref{ZF-spectra}. For temperatures
above \tsdw\ = 138~K a weak Gaussian-Kubo-Toyabe \cite{Hayano79}
like decay  of the muon spin polarization is observed with a
relaxation rate of only 0.095(1)~MHz which is typical for static and
randomly oriented magnetic fields originating from nuclear moments.
This relaxation is also found in superconducting
LaFeAsO$_{1-x}$F$_x$ with $x$=0.10 and 0.075 \cite{Luetkens08}.
Below \tsdw\ a spontaneous muon spin precession with a well defined
frequency is observed which implies a well defined magnetic field at
the muon site. This observation strongly supports our conclusion
from the M\"ossbauer spectroscopy, i.e., an incommensurate spin
density wave as well as a spin glass-like order can be excluded.

\begin{figure}[htbp]
\center{\includegraphics[width=0.95\columnwidth,angle=0,clip]{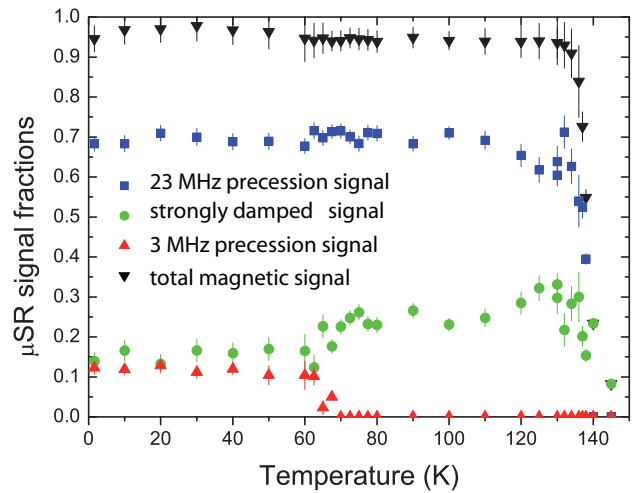}}
\caption[]{Temperature dependence of the $\mu$SR signal fractions
(see text)} \label{Signal_fractions}
\end{figure}

Analyzing the $\mu$SR data in detail proves that 100\% of the sample
volume magnetically orders at \tsdw . A proper description of the
spectra requires at least a two-component relaxation function
indicating magnetically different muon sites. For temperatures above
$\sim$ 70~K a well defined frequency is observed for about 70~\% of
the muons while about 30~\% show a strong relaxation due to a broad
static field distribution. This situation changes below 70~K where a
third component with a low frequency precession ($\approx$ 3 Mhz)
appears which is clearly visible in the raw data at 1.6~K (see
Fig.~\ref{ZF-spectra}). As can be seen from the temperature
dependence of the signal fractions shown in
Fig.~\ref{Signal_fractions}, this signal develops on the cost of the
strongly damped fraction.

The high precession frequency of the first signal implies that these
muons reside very close to the Fe magnetic moments in the FeAs
layers. We associate the weaker signals with more remote  muon
sites, most likely near the LaO layers. All muon sites show the
development of magnetic order below \tsdw\ = 138~K. At present we
have no explanation for the change of the weaker signal below 70~K
but emphasize that it is intrinsic to \lao .

\begin{figure}[htbp]
\center{\includegraphics[width=0.95\columnwidth,angle=0,clip]{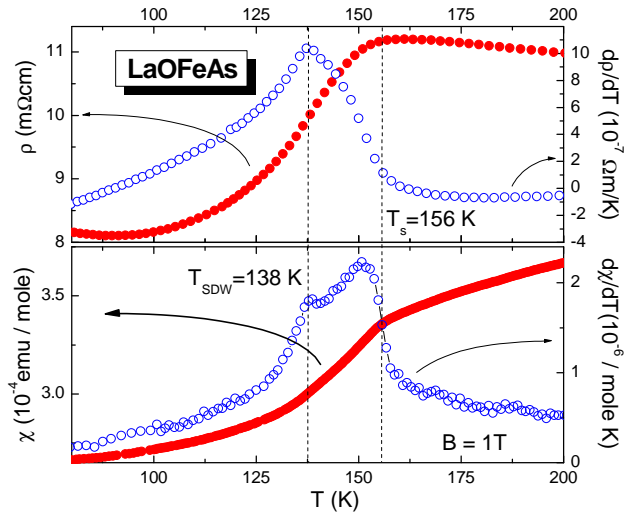}}
    \caption[]{Temperature dependence of (top) the electrical resistivity and (bottom) the magnetisation
    vs. temperature and the respective derivatives $\partial \chi/\partial T$ and $\partial \rho/\partial T$. \ts\ and \tsdw\
    mark the structural phase transition at 156~K and the SDW-formation at 138~K.}
    \label{chi_rho}
\end{figure}

In the literature, evidence for spin density order has been
extracted from magnetization and transport data. Corresponding
measurements for our sample are displayed in figure \ref{chi_rho}.
Both, the electrical resistivity and the magnetization show a clear
anomaly at $\rm T_s~\approx$ 156~K, i.e. well above the magnetic
ordering temperature \tn . Subsequent temperature dependent XRD
studies \cite{Luetkens-phase-diagram-paper} prove that at this
temperature the tetragonal to orthorhombic structural phase
transition occurs which has been observed in recent diffraction
measurements \cite{Cruz08-arXiv,Nomura08}. Anomalies of $\rho$ and
$\chi$ are also found at \tn\ when the temperature derivatives are
considered, as shown in figure \ref{chi_rho}. Both derivatives
exhibit a peak-like anomaly at exactly the same temperature where
the magnetic transition occurs according to our local probe studies.

Recently, the phase transition at \ts\ has been discussed in
connection with the formation of an ''electron nematic phase'' and
broken Ising symmetry \cite{Xu08b-arXiv,KIVELSON08-arXiv}. The
pronounced anomalies of $\rho$ and $\chi$ clearly demonstrate an
intimate coupling between the structural phase transition on the one
hand and the electronic and magnetic properties on the other hand.
At \ts\ the conductivity starts to increase and, surprisingly, this
increase becomes weaker at \tn , i.e. magnetic ordering does not
cause a decrease of electron scattering. The decrease of $\chi$ at
\ts\ shows an enhancement of antiferromagnetic correlations at the
structural phase transition. The anomaly of $\chi$ at \tn\ is
qualitatively very similar, but much weaker. Indeed, analyzing the
magnetic specific heat, which is proportional to $
\partial (\chi \cdot T) /\partial T$, yields a much smaller jump at \tn\
than at \ts . This strong impact of the structural transition on the
magnetism is, however, not visible in our local probe experiments.
At 140~K the $\mu$SR data show only a small volume fraction of
$\approx$ 20 \%  with static local fields, above 145~K we can rule
out any static local magnetic fields. This is typical for a small
distribution of ordering temperatures within the powder sample.
Moreover, there are no signatures of slow spin fluctuations on the
timescale of the Mössbauer and the $\mu$SR experiment above this
temperature.

\begin{figure}[htbp]
\center{\includegraphics[width=0.95\columnwidth,angle=0,clip]{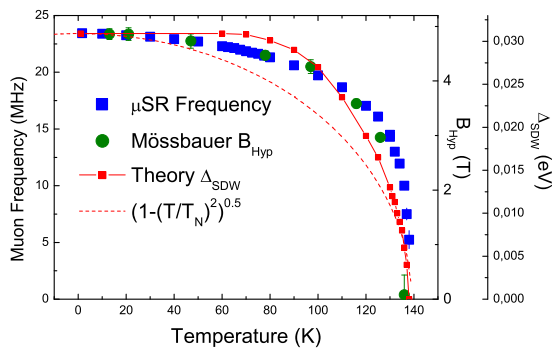}}
\caption[]{Temperature dependence of the main zero field $\mu$SR
frequency, the M\"ossbauer hyperfine field and the calculated SDW
order parameter.} \label{Frequency}
\end{figure}

In order to compare our results with theoretical models we analyse
the temperature dependence of the the magnetic order parameter
obtained from the main muon spin precession frequency and the static
magnetic hyperfine field $B_{hf}$ measured by M\"ossbauer
spectroscopy. As displayed in fig.~\ref{Frequency} both quantities
exhibit the same behavior.

In the following we analyze the experimental data within a
conventional SDW picture. As has been shown previously
\cite{korshunov}, the effective low-energy band structure of the
undoped LaFeAsO can be modeled by a single-electron model
Hamiltonian. For the four-band model considered here the effective
interaction will consist of an on-site Hubbard intraband repulsion
$U$ and the Hund's coupling $J$. There is also an interband Hubbard
repulsion $U'$, which however does not contribute to the RPA
susceptibility. As shown by two of us earlier \cite{korshunov} the
main magnetic instability in the folded Brillouin Zone occurs at the
antiferromagnetic wave vector ${\bf Q}_{AFM}=(\pi,\pi)$ due to the
interband nesting between the hole $\alpha$- and the electron
$\beta$-bands\cite{Zhang08-arXiv,Dong08-arXiv,korshunov,Raghu08-arXiv,Mazin08-arXiv}.
Setting the Hund coupling to $J = 70 $meV and choosing $U=320 $meV
we obtain the ordering temperature $T_N = 138$ K using multiband RPA
susceptibility.

Below \tsdw\ the condition for the SDW instability can be regarded
as a mean-field equation for the SDW order parameter,
$\Delta_{SDW}$. Solving this equation self-consistently we obtain
the temperature dependence of $\Delta_{SDW}(T)$ as shown in figure
\ref{Frequency}. Due to the multi-orbital character of the equation
we find that the SDW gap reaches its saturation rather quickly below
\tsdw\ and the resulting temperature dependence
 deviates from the usual $\sqrt{1-\left(T/T_N\right)^2}$ temperature dependence. In the
experiment a slightly stronger initial increase below \tsdw\ and
further small differences between experiment and theory at lower
temperatures are observed. This may be due to the importance of the
thermal fluctuations neglected here and/or fine details of the band
structure.

From the value $\rm \Delta_{SDW}(T=0K) = 31$ meV we can also
estimate the magnetic moment per Fe site to be $\mu \approx 0.33~
\mu_B$. Considering  that this mean field model does not fully
include quantum fluctuations a value larger than the experimentally
observed one is expected.

In conclusion we present a detailed study of the magnetic order in
\lao\ by means of the local probe techniques M\"ossbauer
spectroscopy and $\mu$SR along with magnetization and resistivity
studies. These experiments prove a commensurate static magnetic
order with a strongly reduced ordered moment of 0.25(5) $\mu_B$ at
the iron site below \tsdw\ = 138~K, well separated from a structural
phase transition at \ts\ = 156~K. A calculation using a four-band
SDW model reproduces the size of the order parameter and the quick
saturation below \tsdw .

\begin{acknowledgments}
We would like to thank M. Zhitomirsky, R. Moessner, and N.M. Plakida
for useful discussions.  We thank M. Deutschmann, S.
M\"uller-Litvanyi, R. M\"uller  and A. K\"ohler for experimental
support in preparation and characterization of the samples. The work
at the IFW Dresden has been supported by the DFG through FOR 538.
\end{acknowledgments}


\end{document}